\documentclass[
twocolumn, preprintnumbers, amsmath,amssymb, superscriptaddress]{revtex4}

\usepackage{graphicx}
\usepackage{dcolumn}
\usepackage{bm}

\newcommand\ee{\end{equation}}
\newcommand\be{\begin{equation}}
\newcommand\eea{\end{eqnarray}}
\newcommand\bea{\begin{eqnarray}}
\newcommand{\sfrac}[2]{{\textstyle\frac{#1}{#2}}}
\newcommand\di{\partial}

\begin{document}




%





\title{Effective field theory for hydrodynamics: \\
thermodynamics, and the derivative expansion}

\author{Sergei Dubovsky}
\email{dubovsky@nyu.edu}

\affiliation{%
Physics Department and Center for Cosmology and Particle Physics,\\
New York University, New York, NY 10003, USA
}%

\author{Lam Hui}
\email{lhui@astro.columbia.edu}

\affiliation{%
Physics Department and Institute for Strings, Cosmology, and Astroparticle Physics,\\
Columbia University, New York, NY 10027, USA
}%

\author{Alberto Nicolis}
\email{nicolis@phys.columbia.edu}

\affiliation{%
Physics Department and Institute for Strings, Cosmology, and Astroparticle Physics,\\
Columbia University, New York, NY 10027, USA
}%

\author{Dam Thanh Son}
\email{son@phys.washington.edu}

\affiliation{Institute for Nuclear Theory, University of Washington,\\
Seattle, WA 98195, USA}

\date{\today}

\begin{abstract}
We consider the low-energy effective field theory describing the infrared dynamics of non-dissipative fluids.
We extend previous work to accommodate conserved charges, and we
clarify the matching between field theory variables and
thermodynamical ones. We discuss the systematics of the derivative
expansion, for which field theory offers a conceptually clear and
technically neat scheme. As an example, we compute the correction to
the sound-wave dispersion relation coming from a sample second-order
term.  This formalism forms the basis for a study of anomalies in
hydrodynamics via effective field theory, which is initiated in a
companion paper.

\end{abstract}

\maketitle


\section{Introduction}
Low energy effective field theory is an extremely powerful tool to 
describe the dynamics of various experimentally accessible physical systems and to parametrize our ignorance of 
short distance physics.  Techniques based on effective field theory are
especially successful when symmetries determine the low energy
particle content and interactions. A classic example of such a
situation is provided by the chiral Lagrangian, describing Goldstone
bosons such as pions.
As a consequence of the symmetries, interaction between Goldstone
bosons are weak at energies small compared to the symmetry breaking
scale, so that the derivative expansion is a natural perturbative
scheme to describe their dynamics.
 
A rich class of phenomena where the low energy degrees of freedom and their 
interactions are also fixed by symmetries to a large extent, is provided by hydrodynamics.  Like the chiral Lagrangian,
hydrodynamics is naturally organized as a derivative expansion,
with the mean free time and the mean free path playing a role
similar to the symmetry breaking scale
for the chiral Lagrangian.
The similarity with the pion chiral Lagrangian goes even further.  
Indeed, hydrodynamical degrees of freedom are actually Goldstone modes,
either of space translations spontaneously broken by the presence of the medium (phonons) \cite{DGNR}, or of global conserved charges carried by the fluid \cite{Son}. 
  
Nevertheless, the traditional approach to hydrodynamics is quite
different from conventional effective field theory. One starts with a
set of conservation law for a set of ``composite" objects---the
energy-momentum tensor and the conserved currents.  The derivative
expansion enters at the level of the ``constitutive relations,'' which
express the composite objects in terms of more elementary fluid
quantities---the fluid velocity field and local thermodynamical
variables.  As discussed in more detail below, this prescription
becomes somewhat ambiguous as one goes to higher orders in the
derivative expansions.

The main goal of this paper is to demonstrate that, at least for a
fluid without dissipation, hydrodynamics can be recast into the
traditional effective field theory language.  Namely, just like for
the pion chiral Lagrangian, we start with a set of Goldstone fields
determined by the symmetries of the fluid. Then we write the most
general effective action compatible with the symmetries, and make use
of the conventional derivative expansion employed in effective field
theories.  The main non-trivial step in this program is identifying
the relevant symmetries; once it is done, the rest is automatic.
However, the translation from field theory to the conventional
language of hydrodynamics requires more work, and can become quite
laborious as one goes to higher orders in the derivative expansion.
To a large extent, this is related to the aforementioned subtleties in
the traditional hydrodynamical derivative expansion at higher orders.

Our viewpoint here is that for many purposes it is convenient and
instructive to consider effective field theory as a {\it definition}
of what dissipationless hydrodynamics is.  As illustrated in the main
text, it is straightforward then to calculate various physical
effects, such as higher derivative corrections to the sound wave
dispersion relation. To find what a particular field theory operator
corresponds to in the traditional language of constitutive relations
may be more challenging (technically, not conceptually), but also not
really necessary.

The organization of this paper is as follows.  In
section~\ref{sec:generalfluids} we identify the symmetries relevant
for describing a general perfect fluid at non-zero temperature and
chemical potential.  Here we build on the previous results of
Refs.~\cite{Son} and \cite{DGNR}, where the effective field theory formalism
for zero temperature superfluids and barotropic normal fluids was
developed.  In section~\ref{sec:thermodynamics} we establish the
dictionary between field theory and conventional thermodynamical
variables at the leading order in the derivative expansion.

We continue in section~\ref{sec:IS} by extending the dictionary
between field theory and hydrodynamics to higher orders in the
derivative expansion, following the standard procedure (see, e.g.,
Ref.~\cite{IS}). We exemplify this prescription by working out
explicitly several simple examples.  First, we consider
one-derivative corrections on the field theory side.  From
hydrodynamics, one does not expect any non-dissipative terms at this
order. In agreement with this expectation, we show that on the field
theory side these corrections can be shifted to higher orders by a
field redefinition.  As a cross-check of our prescription, in the
Appendix we show that the same procedure of field redefinition can
be performed in the hydrodynamic theory.
(In section \ref{redundant} we prove
that the same property holds at all orders in the derivative
expansion: if a Lagrangian term can be removed via a field
redefinition, it has no effects on the hydrodynamics of the system.)
We also calculate the
correction to the sound wave dispersion relation following from a
sample nontrivial two-derivative term.

We conclude in section~\ref{conclusions} by mentioning a number of possible applications of our formalism, some of which we are already investigating in detail.

Our main
emphasis here is not on the reformulation of the 
hydrodynamical equations as an action principle, but rather on
the underlying principles behind this construction: symmetry and symmetry
breaking, Goldstone bosons, and derivative expansion.
Apart from Refs.~\cite{Son,DGNR} mentioned above, earlier works similar in
spirit to the present paper include
\cite{Greiter:1989qb,Leutwyler:1996er,Son:2005ak,ENRW}.  
Our approach may be useful for ``holographic fluids" where---as 
emphasized in Ref.~\cite{Nickel:2010pr}---at low energies the Goldstone
dynamics can be parameterized without any reference to the microscopic
theory.

There has been recent interest in the hydrodynamics of systems
carrying anomalous charges, starting with Ref.~\cite{SS}.  Here, we
restrict ourselves to fluids carrying ordinary conserved charges, and
we avoid the subtleties associated with the presence of quantum
anomalies. We are devoting a companion paper precisely to those
subtleties, and to the resulting interesting effects \cite{paper2}.

Finally, we will deal directly with {\em relativistic} hydrodynamics,
i.e., in our field theory we will impose (spontaneously broken)
Lorentz rather than Galilei invariance. 
This choice makes the treatment somewhat simpler.  The
non-relativistic limit can be taken at any stage in our analysis, if
needed. 


\section{Fluids with conserved charges}
\label{sec:generalfluids}
Consider a perfect fluid in $d$ spatial dimensions.  Its low-energy 
degrees of freedom can be chosen to be $d$ scalar fields
\be
\phi^I = \phi^I(\vec x, t) \qquad I = 1, \dots, d \; ,
\ee
giving the comoving (Lagrangian) coordinates of the volume element
occupying physical (Eulerian) position $\vec x$ at time $t$. This
description is reviewed extensively in Refs.~\cite{DGNR, ENRW}, to
which we refer the reader for details \footnote{Earlier works adopting the same parameterization for the fluid degrees of freedom include, e.g.,  Refs.~\cite{carter, CL}.}.  There is an inherent
arbitrariness in labeling the volume elements via comoving
coordinates. It can be fixed, for instance, by choosing these to be
aligned with the physical ones when the fluid is in equilibrium at
some reference external pressure,
\be \label{phi=x}
\phi^I = x^I \qquad \mbox{(equilibrium)} \; .
\ee
With this choice of  field variables, the fluid's dynamics must enjoy the 
internal symmetries \cite{DGNR}
\begin{align}
\phi^I & \to  \phi^I + a^I \; ,  \qquad a^I = {\rm const}   \label{shift}\\
\phi^I & \to  R^I {}_J \, \phi^J  \; , \qquad  R \in SO(d)   \label{rotate} \\
\phi^I & \to  \xi^I (\phi) \;, \qquad \det \big( \di \xi^I / \di \phi^J \big)= 1
    \label{diff}
\end{align}
on top of $(d+1)$-dimensional Poincar\'e invariance. In particular,
eq.~\eqref{diff} corresponds to the fluid's insensitivity to (static)
non-compressional deformations.  Note that even though all small
perturbations about \eqref{phi=x} are allowed, non-perturbatively
every field configuration must define a time-dependent diffeomorphism
between physical and comoving space. That is, at any given time it
must be an invertible function of $\vec x$ \footnote{To avoid
confusion, let us stress that this condition does not imply that our
variational problem is constrained. Locally the condition of
invertibility is simply the {\it in}-equality $\det (\di\phi/\di x)\neq
0$ and does not lead to any constraints on the variations of
$\phi$.}.

Suppose now that the fluid carries a conserved charge. There should be an associated $U(1)$ symmetry in our field theory. It does not seem sensible 
to realize this symmetry using  the $\phi^I$ fields only: they represent
the comoving coordinates and physically cannot transform under a particle
number symmetry.  Besides, the $\phi^I$'s are non-compact.  
We should augment the field content to represent the particle number symmetry.
The most economical addition is a real {\em phase} $\psi(\vec x,t)$ that shifts under it:
\be \label{U(1)}
U(1): \quad \psi \to \psi + c \; .
\ee

We are thus led to construct the low-energy effective field theory for 
the $\phi^I$'s and $\psi$, subject to the symmetries 
\eqref{shift}--\eqref{U(1)} and to Poincar\'e invariance.
However there should be an additional constraint on the theory, as we know
that in ordinary perfect-fluid hydrodynamics, the particle number current 
is comoving with the fluid:
\be \label{j}
j^\mu = n u^\mu \; .
\ee
(The fluid velocity $u^\mu$  can be defined through, e.g., the 
entropy current of the fluid.)
This guarantees that, in normal fluids, sound waves are the only
propagating wave solutions: the new charge
degree of freedom does not introduce new waves (in contrast to superfluids
where there are first and second sounds).
From our discussion so far, it is not obvious how this is going to
arise. It turns out that enforcing eq.~\eqref{j} is equivalent to imposing a
{\em new} symmetry. If charge flows with the fluid, charge
conservation is obeyed separately within each comoving fluid
element (recall that diffusion is a dissipative effect and is outside the scope of our theory). 
This means that charge conservation is not
affected by an arbitrary comoving position-dependent redefinition of
the charge units. In other words, the $U(1)$ symmetry \eqref{U(1)} can
be made comoving position-dependent:
\be \label{f(phi)}
\psi \to \psi + f(\phi^I) \; ,
\ee
where $f$ is a generic function. We will see below that this is exactly what we need to enforce \eqref{j} in our field theory.
Since the chemical potential will turn out to be the simplest invariant under this symmetry, and for lack of a better term, we dub this new symmetry `chemical shift'.

Note that beyond the leading order in the derivative expansion, the
particle number current is not necessarily parallel to the fluid flow. 
Nevertheless, from the
field theory viewpoint, it is natural to impose the symmetry
(\ref{f(phi)}) to all orders. This automatically implies the existence
of a quantity conserved along the flow. This property is not obvious
in the conventional hydrodynamic language, but also the very notion of
dissipationless fluid may be ambiguous there beyond the leading
order. Our viewpoint is that the effective field theory
characterized by the symmetries \eqref{shift}--\eqref{U(1)} and
(\ref{f(phi)}) provides a natural and unambiguous definition of
dissipationless (nonanomalous) hydrodynamics to all orders.

We are thus looking for the most general relativistic Lagrangian that is invariant under \eqref{shift}--\eqref{U(1)} and under \eqref{f(phi)}. At low energies it should be organized as a derivative expansion. At lowest order, we should have one derivative per field, because of \eqref{shift} and \eqref{U(1)}:
\be
{\cal L} = {\cal L}(\di \phi^I, \di \psi)
\ee
Because of \eqref{rotate} and \eqref{diff}, the $\di \phi^I$'s should enter in the combination
\cite{DGNR, ENRW}
\begin{align}
J^\mu & \equiv  \epsilon^{\mu\alpha_1 \dots \alpha_d} \, 
  \di_{\alpha_1} \phi^1 \dots \di_{\alpha_d} \phi^d \\
& = \frac{1}{d!} \epsilon^{\mu\alpha_1 \dots \alpha_d} \, 
  \epsilon^{I_1 \dots I_d} 
\, \di_{\alpha_1} \phi^{I_1} \dots \di_{\alpha_d} \phi^{I_d} \, . \label{J}
\end{align}
(we define the $(d+1)$-dimensional $\epsilon$ tensor by
$\epsilon^{01\dots d} = +1$.) 
$J^\mu$ has an important property:
it is a vector field along which the comoving coordinates do not change:
\be \label{comoving}
J^\mu \di_\mu \phi^I = 0 \;, \qquad I =1, \dots, d \; .
\ee
Thus it is natural to define the fluid's four-velocity as a unit
vector aligned with $J^\mu$:
\be \label{u}
u^\mu = \frac{1}{b} J^\mu \; , \quad
b \equiv \sqrt{-J_\mu J^\mu} \; .
\ee
We use the `mostly plus' signature.
From a geometric viewpoint, $J^\mu$ is the current of fluid points \cite{DGNR}.

Now we make use of the chemical shift \eqref{f(phi)}. We notice that,  because of \eqref{comoving}, the combination
\be
J^\mu \di_\mu \psi
\ee
is invariant. In fact, it is the only invariant at this order in derivatives. The reason is that the only vector that is orthogonal to $\di_\mu f$ for generic $f(\phi^I)$, is the `vector product' of all the $\di \phi^I$'s, eq.~\eqref{J}.

To summarize, at lowest order in the derivative expansion,
the Lagrangian can depend on the $\phi^I$'s and on $\psi$ through $J^\mu$ and $J^\mu \di_\mu \psi$ only. It must be a Poincar\'e scalar. We choose to parametrize it as
\be \label{F(b,y)}
S = \int\! d^4x \, F(b,y) \; ,
\ee
where $b$ is given above, $y$ is defined as
\be
y \equiv u^\mu \di_\mu \psi = \frac1b \, J^\mu \di_\mu \psi \; ,
\ee
and $F$ is a generic function. We will see below that $F$ is related to the equation of state of our fluid.

The same field theoretical description of conserved charges in hydrodynamics---with emphasis on the same symmetry \eqref{f(phi)}---has been worked out independently by Sibiryakov \cite{sergey}.


The Noether current associated with $\psi$'s shift symmetry \eqref{U(1)} is
\be \label{noether}
j^\mu = \frac{\partial F}{\partial y} u^\mu \equiv F_y u^\mu  ,
\ee
and is
indeed comoving with the fluid, as desired. 
The chemical shift (\ref{f(phi)}) is an infinite-dimensional symmetry, and as a consequence there are infinitely many currents associated with it. They are
\be
j^\mu_{(f)} = F_y \, f(\phi^I) \,  u^\mu  = f(\phi^I) \, j^\mu\; .
\ee
Their conservation is implied by that of $j^\mu$:
\be
\di_\mu j^\mu_{(f)} = f(\phi^I) \di_\mu j^\mu + F_y \, \di_\mu f(\phi^I) u^\mu \; ;
\ee
the second term vanishes identically, thanks to \eqref{comoving}.

To develop more intuition on why (\ref{F(b,y)}) is the correct 
description of a fluid carrying a conserved charge, let us consider how the system behaves in the presence of an external gauge field $A_\mu$.
In the field theory, the natural way to describe this is  to gauge the $\psi$ shift symmetry, {\it i.e.} to replace $\di_\mu\psi\to\di_\mu\psi+A_\mu$ everywhere in the action (\ref{F(b,y)}). The resulting action has a non-linear dependence on the gauge field. This may appear puzzling from the hydrodynamical point of view, where on physical grounds one may expect the following linear coupling between the gauge field and the fluid
 \bea
S_{\rm int} = \int\! d^4 x \, N(\phi^I) A_\mu J^\mu 
		=\int\! d^3 \phi \, d \tau \, N(\phi^I) u^\mu A_\mu
		 \,  \label{Sint}
\eea
where $N(\phi^I)$  is the comoving charge density, and at the
last step we switched to the comoving frame, {\it i.e.} chose $\phi^I$
as space coordinates and used $\vec x(\phi,\tau)$ as the dynamical
variables, with $\tau$ denoting the proper time along the comoving worldline.
This switch is performed by making use of the identity \cite{DGNR}
\be
d^3 \phi \, d \tau = \left| \det \frac{\di \phi^I}{ \di x^i} \right| d^3 x   \, \sqrt{1-\dot {\vec x} \, ^2} \, dt =  b  \, d^4 x\;.
\ee 

To see the relation between the two descriptions let us rewrite also the (gauged) fluid action (\ref{F(b,y)}) in  comoving coordinates,
\be
\label{Scom}
S=\int\! d^3 \phi \, d \tau \, b^{-1} F(b,\di_\tau\psi+u^\mu A_\mu)
\ee
The action (\ref{Scom}) depends on $\psi$ only through its time derivative, so that the canonically conjugate momentum
\[
\Pi_\psi=b^{-1}F_y
\]
is time independent on the classical solutions,
$\Pi_\psi=\Pi_\psi(\phi^I)$.  By making the Legendre transform
w.r.t.~$\di_\tau \psi$ ({\it i.e.}, switching to the Routhian description
\footnote{Given a system with Lagrangian ${\cal L}(\dot q, ...)$,
where the ellipsis denotes other fields, the fact that the
conjugate momentum ($p$) to $q$ is conserved can be used to integrate out
$q$. The dynamics is then described by the effective Lagrangian
${\cal L}_{\rm eff} = {\cal L} - p \dot q$, where all $\dot q$
dependence should be eliminated using $p(\dot q, ...) = {\rm
  constant}$ \cite{LL}. In our example, $\psi$ plays the role of $q$. Thus
$- p \dot q \rightarrow - \Pi_\psi \partial_\tau \psi = 
- \Pi_\psi (y - u \cdot A)$.})
we arrive at the classically equivalent action,
\be
\label{Routhian}
 \int\! d^3 \phi \, d \tau \left[ b^{-1} F(b,y(b,\Pi_\psi))-\Pi_\psi y(b,\Pi_\psi)+\Pi_\psi u^\mu A_\mu\right] \;,
\ee
with a linear dependence on $A_\mu$. So, the two descriptions are
indeed equivalent upon the
identification $N(\phi)=\Pi_\psi(\phi)$, which is consistent with the field theory/thermodynamics dictionary we are now going to establish.

\section{Thermodynamics}
\label{sec:thermodynamics}
We now want to make contact with 
the standard hydrodynamic and thermodynamic description of a fluid carrying 
a conserved charge. 
From the action
\begin{align}
S & = \int\! d^{d+1} x \,  F(b, y)
\end{align}
we can derive the stress energy tensor by varying with respect to the metric. We get
\be
T_{\mu\nu} = (F_y y - F_b b ) B^{-1}_{IJ} \di_\mu \phi^I \di_\nu \phi^J + (F - F_y y) \eta_{\mu\nu} \; ,
\ee
where $F_b$ is the $b$-derivative of $F$, and  the matrix $B^{IJ}$ is defined as 
\be
B^{IJ} \equiv  \di_\mu \phi^I \di^\mu \phi^J  \; . 
\ee
From the definition of $J^\mu$ it is straightforward to
see that
\be
\det B^{IJ} = b^2
\ee
---which we used to derive $T_{\mu\nu}$---and
\be \label{Pmn}
B^{-1}_{IJ} \di_\mu \phi^I \di_\nu \phi^J = \eta_{\mu\nu} + u_\mu u_\nu \equiv P_{\mu\nu}\; .
\ee
We can thus rewrite the stress-energy tensor in a more familiar form:
\be
T_{\mu\nu} = (F_y y-F_b b) \, u_\mu u_\nu + (F - F_b b) \eta_{\mu\nu}
\ee
The fluid's energy density and pressure therefore are:
\be \label{rhoandp}
\rho = F_y y - F \; , \qquad p = F - F_b b \; .
\ee
Likewise, by comparing (\ref{noether}) and (\ref{j}) we get that the fluid's charge density is 
\be
n = F_y \; .
\ee

We can get the chemical potential $\mu$, the entropy density $s$, and the temperature $T$ by imposing the thermodynamics identities
\be \label{thermoidentities} 
\rho + p = T s + \mu n \; , \qquad d \rho = T \, ds + \mu \, dn \; .
\ee
Before doing so, it is worth pointing out that our vector $J^\mu$ is an {\em identically} conserved current:
\be
\di_\mu J^\mu = 0 \qquad \mbox{(identity)} \; .
\ee
This follows straightforwardly from its definition. Moreover, as we already mentioned, it is aligned with the fluid's four velocity, $J^\mu = b  \, u^\mu$. These two properties invite identifying $J^\mu$ with the {\em entropy} current, and $b$ with the entropy density,
\be \label{s=b}
s = b \; .
\ee
That entropy is conserved identically---i.e., `off-shell'---in our
non-dissipative field theory, makes perfect sense.  We could imagine
coupling our field theory Lagrangian to external sources. These
sources could perform work on the system, but would not exchange heat
with it. In such an instance our fluid would be `off-shell', but
entropy would still be conserved~\footnote{Of course, the same would
  be true if the entropy current were a Noether current, provided the
  coupling to external sources preserve the corresponding
  symmetry. The difference is that the entropy current would depend on
  the sources in this case.  A priori there is nothing wrong with
  this, and this may lead to an alternative dictionary between field
  theory and hydrodynamics. This ambiguity may be related to the
  ``integration constants" of anomalous hydrodynamics~\cite{paper2}.}.
The interpretation of $b$ as entropy density has been derived independently in \cite{sergey}.

From the first identity in eq.~\eqref{thermoidentities} , we thus get
\be
T = -F_b \; , \qquad \mu = y \; , 
\ee
which is consistent with the second identity too. (For the differential identity, one should express $d\rho$ in terms of $db$ and $dF_y = dn$.) There is of course an ambiguity in the overall normalization of $s$ and $T$---we could multiply $s$ and divide $T$ by the same constant, without affecting the thermodynamical identities. This is of course related to Boltzmann's constant, which does nothing but defining the units of temperature.

In conclusion, our Lagrangian $F$ \eqref{F(b,y)} is naturally a function
of the entropy density $b$ and of the chemical potential $y$. 
It can be thought of as a somewhat
unusual thermodynamic potential: $dF = -T ds + n \, d\mu$.
It is related to the equation of state $\rho(s,n)$ or $p(T, \mu)$ via either
of the Legendre transforms in (\ref{rhoandp}).


\section{Higher derivative corrections}
\label{sec:IS}
Hydrodynamics is naturally organized as a derivative expansion.  For instance, for the hydrodynamic regime of a weakly coupled system of particles, the natural expansion parameters are the fields' time derivatives times the mean free time, and the fields' spacial gradients times the mean free path. The standard hydrodynamical and thermodynamical variables $\rho$, $p$, $u^\mu$, etc., correspond in our field theory to `composite operators' involving one derivative per $\phi^I$ or $\psi$ field. To reproduce higher-order corrections to the perfect fluid hydrodynamics, involving derivatives of such variables, we need to include in our field theory Lagrangian terms involving correspondingly more derivatives than the lowest order ones. Because of this, we will number higher derivative corrections starting from our lowest order Lagrangian. That is, when we talk about `one-derivative terms', we mean terms involving overall one more derivative than one per field; and so on. Of course,
  since we will work at the level of the action, our field theory will be conservative by construction. That is, our approach will not be able to reproduce dissipative effects.

We can adapt to our field-theoretical framework the procedure of
dealing with higher-derivative hydrodynamics outlined by Israel and
Stewart (IS) \cite{IS}. The question is---essentially---how to apply
thermodynamics to a fluid in the presence of spacial gradients and
time derivatives, which typically signal that the system is not in
complete equilibrium, even though there is some form of
local thermodynamic equilibrium.
IS argue that one should proceed as follows. At any given spacetime point $x$, the stress-energy tensor $T_{\mu\nu}(x)$ and the charge current $j_\mu(x)$ are perfectly well defined quantities, even for out-of-equilibrium systems. 
For us, they descend straightforwardly from the Lagrangian, respectively  by varying with respect to the metric and as the Noether current associated with the shift symmetry \eqref{U(1)}.
One defines the local energy density $\rho(x)$ and the local charge density $n(x)$ by taking contractions with the local $u^\mu(x)$:
\be
\rho \equiv  u^\mu u^\nu T_{\mu\nu} \; \qquad n \equiv - u^\mu j_\mu 
\ee
($u^\mu$ is time-like---hence the minus sign.) In the presence of gradients, the fluid velocity field $u^\mu$ is itself ambiguous. For instance, the energy flow and the charge flow are typically not aligned with each other. Such an ambiguity is harmless \cite{IS}, and in fact, typically it can be used to simplify some algebra. Moreover, $\rho$ and $n$ are particularly well-behaved from this viewpoint, since they are unaffected by small variations of $u^\mu$, at first order in these variations. One then defines the local values of all other thermodynamic variables by applying the {\em equilibrium} equation of state to the local $\rho$ and $n$ thus defined:
\be
p(x) = p_0(\rho, n) \; , \qquad \mu(x) = \mu_0(\rho, n) \;, \qquad \mbox{etc.}
\ee
The subscript zeroes are there no remind us that we should use precisely the same functions of $\rho$ and $n$ as for the fluid in equilibrium, i.e., in the absence of gradients. Finally, one goes back to the stress-energy tensor and the current, subtracts the perfect fluid part according to the above identifications, and interprets whatever is left as the higher-derivative corrections:
\begin{align}
T_{\mu\nu} & = (\rho + p)u_\mu u_\nu + p \, \eta_{\mu\nu} + \delta T_{\mu\nu} \\
j_\mu & = n \,u_\mu + \delta j_\mu 
\end{align}
 
The IS prescription is in a sense merely a convenient {\em definition} of what we might want to mean by thermodynamical quantities for an out-of-equilibrium fluid. It has the advantage of establishing an unambiguous dictionary. Moreover, at first order in gradients all such quantities  are independent of the precise choice of $u^\mu$, thus appearing perfectly well-defined, and physical. On the other hand, at second order and up they all become inherently $u^\mu$-dependent, and attaching any precise physical meaning to them becomes more and more dubious \cite{IS}. As an important example for us, the entropy density defined as above,
\be
s \equiv s_0 (\rho, n)
\ee
will not coincide in general with the ``observed'' entropy density $-u_\mu s^\mu$, where $s^\mu$ is the entropy current \cite{IS}.  This is well defined even for (slightly) out-of-equilibrium systems, because the second law has to hold for them: $\di_\mu s^\mu \ge 0 $.

As we already emphasized, for us there is no entropy production, even off-shell---because our system is conservative, by construction---and the entropy current is naturally identified with the {\em identically} conserved current \eqref{J}:
\be
s^\mu \equiv J^\mu \; .
\ee
$J^\mu$ also defines an unambiguous rest frame for the fluid, which differs in general from those associated with the $U(1)$ current $j^\mu$ and with the stress-energy tensor. We will refer to this frame as the `field-theory frame' or the `entropy frame'. 

We want to stress that our field theory can be taken as an independent {\em definition} of a non-dissipative fluid with mild gradients, that is, mildly out of equilibrium. Its thermodynamics may be ambiguous---{\em \`a la} Israel and Stewart---but its dynamics are not. For non-thermodynamical questions, thermodynamics and the IS procedure can be bypassed entirely and the relevant observables can be computed directly from the field theoretical description.
For instance, this is the case for the higher-derivative corrections to the sound-wave dispersion law of sect.~\ref{secondorder}.

We start by considering one-derivative terms in our field theory.
As is well known, absent anomalies \cite{SS}, the only one-derivative corrections one can write down for hydrodynamics are dissipative---they are associated with shear viscosity, bulk viscosity, and conductivity \cite{LLfluid}. As such, they cannot be reproduced by our field theory. We thus expect that---once interpreted correctly---one-derivative corrections to our Lagrangian will be trivial. As a warmup, we consider one-derivative corrections to the dynamics of a zero-temperature superfluid, which also turn out to be trivial.

\subsection{First-order superfluid dynamics}\label{firstordersuper}

Consider a relativistic superfluid at zero temperature. Its zeroth-order field theoretical description has been worked out in \cite{Son}. It involves a scalar field $\psi$ with a shift symmetry $\psi \to \psi + a$, with a time-dependent vev $\langle \psi(x) \rangle \propto t$. The lowest order Lagrangian is
\be \label{P(X)}
{\cal L}_0 = P(X) \;, \qquad X \equiv (\di \psi)^2 \; . 
\ee
The associated stress-energy tensor and current are
\begin{align}
T^0_{\mu\nu} & = -2 P'(X) \di_\mu \psi \di_\nu \psi + P \, \eta_{\mu\nu} \\
j^0_\mu & = 2 P'(X) \di_\mu \psi \; . 
\end{align}
At this order it is natural to define $u_\mu$ along $\di_\mu \psi$:
\be
u^0_\mu \equiv -\frac{\di_\mu \psi}{\sqrt{-X}} \; . 
\ee
The minus sign upfront is to some extent a matter of convention. We are assuming that the superfluid's ground state has $\psi \propto + t$.
Taking the appropriate contractions with $T_0^{\mu\nu}$ and $j^\mu_0$ we get
\be \label{rhoandn}
\rho_0 = 2 P'  X - P \; , \qquad n_0 = -2 P' \sqrt{-X} \; , \qquad p_0 = P \; .
\ee
These are consistent with the zero-temperature thermodynamic identities
\be
\rho + p = \mu n \;, \qquad d \rho  = \mu \, dn \; ,
\ee
with chemical potential
\be \label{mu}
\mu_0 =  \sqrt{-X} \;.
\ee
The function $P(X)$ is thus naturally interpreted as the equation of state, giving the pressure as a function of the chemical potential \cite{Son}:
\be \label{p(mu)}
p_0 = P (- \mu_0^2 ) \; .
\ee

We now add all possible one-derivative corrections consistent with the symmetries. They take the form
\be \label{onederivativesuper}
\Delta {\cal L} = G(X) \di^\mu \psi \di_\mu X \; . 
\ee
There is another possible structure at this order---$H(X) \Box \psi$---which however can be rewritten as above upon integrating by parts. This just redefines $G(X)$, which is arbitrary anyway.
Before  proceeding with the IS prescription, we notice that $\Delta {\cal L}$ can be removed by a field redefinition, at the price of introducing {\em higher} derivative corrections, with two or more derivatives. The reason is that it vanishes on the zeroth-order equations of motion, {\em for any $G(X)$}. Indeed:
\begin{align}
\Delta {\cal L} & = \frac{G(X)}{P'(X)} \di_\mu X \,  P'(X) \di^\mu \psi \\
& = \di_\mu \tilde G(X) \, P'(X) \di^\mu \psi \; ,
\end{align}
where $\tilde G \equiv \int G/P' dX$. If we integrate by parts, we get precisely the equations of motion associated with the zeroth order Lagrangian \eqref{P(X)}. As a result $\Delta {\cal L}$ can be removed by redefining $\psi$:
\be \label{psi'}
\psi = \psi' - \sfrac12 \tilde G(X') \; .
\ee
As usual, as a byproduct of this redefinition we get higher order terms, with two or more derivatives.
The effects associated with $\Delta {\cal L}$ can thus be deferred to higher orders in the derivative expansion.
Whenever  something like this happens, the corresponding Lagrangian term is said to be `redundant'.
Not surprisingly, if one applies the IS procedure to a redundant term {\em before} performing the field redefinition that removes it, one also gets trivial effects at the order under consideration. We prove this  in sect.~\ref{redundant} in broad generality, and in the Appendix for the case under consideration.

\subsection{First-order fluid dynamics}\label{firstorderfluid}

For our fluid, the most general first order terms 
consistent with our symmetries are
\be \label{onederivativefluid}
\Delta {\cal L} = f_1(b,y) \, J^\mu \di_\mu b +  f_2(b,y) \, J^\mu \di_\mu y \; ,
\ee
where $f_1$ and $f_2$ are generic functions. There is in principle another one-derivative structure,
\be
f_3(b,y) \, J^\nu J^\mu \di_\mu J_\nu \; ,
\ee
but this is in fact of the same form as the first term in \eqref{onederivativefluid}, as $J^\nu \di_\mu J_\nu= - \sfrac12 \di_\mu b^2$.
We now show that $\Delta {\cal L} $ is redundant, and it can thus be removed via a  field redefinition. To this end, it is useful to inspect the zeroth order equation of motion for $\psi$. It is the conservation of $j^\mu$:
\be
\di_\mu (F_y u^\mu ) = 0 \; .
\ee
Given that  $J^\mu$ is identically conserved, it can be pulled out of the derivative. One gets
\be \label{n/s}
J^\mu \di_\mu N = 0 \; , \qquad N \equiv F_y/b \; .
\ee
$N$ is the inverse `entropy per particle', and the above equation states the well-known fact that---at zeroth order---such a quantity is conserved along the flow. $N$ is of course a function of $b$ and $y$. It is useful to change variables in \eqref{onederivativefluid} and express everything in terms of $b$ and $N$:
\be
\Delta {\cal L} = \tilde f_1(b,N) \, J^\mu \di_\mu b +  \tilde f_2(b,N) \, J^\mu \di_\mu N \; .
\ee
The second term vanishes on the zeroth order equation of motion \eqref{n/s}. The first does not, but we can get rid of it via the following trick. We define a new function $g(b,N)$ such that
\be
\tilde f_1(b,N) = \di_b g(b,N) \; ,
\ee
that is, $g(b,N)  \equiv \int \tilde f_1(b,N) \, db $. We thus get
\be
\Delta {\cal L} = J^\mu \di_\mu g(b,N) +  \tilde f_3(b,N) \, J^\mu \di_\mu N  \; ,
\ee
where $\tilde f_3 \equiv \tilde f_2 - \di_N g$.
The first term now is a total derivative---because $J^\mu$ is identically conserved---while the second vanishes on the zeroth-order field equations. As a result, all physical effects associated with $\Delta {\cal L}$ can be moved via a field redefinition to higher orders in the derivative expansion. As we mentioned, this is consistent with the absence of non-dissipative one-derivative corrections to hydrodynamics.

\subsection{Sample second order correction}\label{secondorder}

We leave for future work a thorough study of second order corrections in our field theory. Here instead, as an illustration, we apply the IS procedure outlined above to a sample two-derivative term allowed by all symmetries:
\be \label{db2}
\Delta {\cal L} = \alpha \, \di_\mu b \, \di^\mu b \; ,
\ee
where $\alpha$ is an arbitrary coupling constant. Also, for simplicity we assume that our fluid does not carry any conserved charge, in which case the lowest order Lagrangian does not depend on our charge field $\psi$,
\be
{\cal L}_0 = F(b) \; .
\ee
The contribution to the stress-energy tensor associated with $\Delta {\cal L}$ is
\be
\Delta T_{\mu\nu} = -2\alpha \, \di_\mu b\, \di_\nu b+ 2\alpha \,   b \Box b P_{\mu\nu}  \; ,
\ee
where $P_{\mu\nu}$ is the transverse projector of eq.~\eqref{Pmn}. As  $u^\mu$, we will still use our zeroth-order definition \eqref{u}---i.e.~that associated with the entropy current $J^\mu$. 
The correction to the energy density coming from $\Delta {\cal L}$ therefore is
\be \label{Deltarho2}
\Delta \rho  \equiv  u^\mu u^\nu \, \Delta T_{\mu\nu} = -2\alpha (u\cdot \di b)^2 \;.
\ee
We are now supposed to derive the other thermodynamic variables via the {\em zeroth order} relations between $\rho$, $s$, and $p$, which define the equilibrium equation of state. We find it convenient to express everything in terms of the entropy density. From eqs.~\eqref{s=b} and \eqref{rhoandp}, and setting $F_y \to 0$ we get
\begin{align}
\rho & = \rho_0 - F'(s_0) \Delta s  \\
p & = p_0 - s_0 F''(s_0) \Delta s \; ,
\end{align}
where $s_0 \equiv b$ is the zeroth-order entropy density.
A comparison with \eqref{Deltarho2} gives immediately
\be
\Delta s =\frac{1}{F'} 2\alpha (u \cdot \di b)^2
\ee
and therefore
\be
\Delta p = - \frac{b F''}{F'} 2\alpha (u \cdot \di b)^2
\ee
Notice that the combination $b F''/F'$ is precisely the squared speed of sound \cite{DGNR, ENRW}, so that $\Delta p = c_s^2 \Delta \rho$, as implied by our using the zeroth-order equation of state.

We can now rewrite the full stress energy tensor in terms of the `physical' quantities $\rho$, $p$, $s$ redefined as above. We get
\begin{align} 
T_{\mu\nu} & \equiv  T^0_{\mu\nu}  + \Delta T_{\mu\nu} \nonumber \\
& = (\rho+p) u_\mu u_\nu + p \, \eta_{\mu\nu} \nonumber \\
& +2 \alpha P_{\mu\nu} (s \Box s +2 c_s^2 (u \cdot \di s)^2) \nonumber \\
& + \alpha P_{\mu \alpha} P_{\nu \beta} \, \di^\alpha s \, \di^\beta s \nonumber \\
& - 2 \alpha (u \cdot \di s) \, \di^\alpha s \, P_{\alpha (\mu} u_{\nu)} \; .  \label{fullTdb2}
\end{align}

The gap in simplicity and clarity between our field theoretical starting point---eq.~\eqref{db2}---and the more standard hydrodynamical parametrization of the same second-order correction---eq.~\eqref{fullTdb2}---is manifest.
For instance, if we expand eq.~\eqref{db2} in small perturbations about a homogeneous and static background,
\be
\phi^I = x^I + \pi^I \; ,
\ee
we get directly a correction to the quadratic Lagrangian for the sound waves \cite{DGNR, ENRW}
\begin{align}
{\cal L}_0 + \Delta {\cal L} & \to \sfrac12 (\rho+p) \big[ \dot {\vec \pi}^2 - c_s^2 \big(\vec \nabla \cdot \vec \pi\big)^2 \big] \\
& + \alpha  (\di_\mu \vec \nabla  \cdot \vec \pi) ^2 \; .
\end{align}
At low momenta, this corresponds to a quartic correction to the dispersion law:
\be
\omega^2 \simeq c^2_s \, k^2 + 2 \alpha\sfrac{1-c_s^2}{(\rho+p)} \, k^4 \; .
\ee

Finally, notice what we anticipated above: that the entropy density defined following the IS prescription does not coincide, at second order, with that associated with the entropy current:
\be
s = b + \Delta s \neq -u_\mu J^\mu = b \; .
\ee


\section{Redundant couplings}\label{redundant}

As we saw in the last section, at the level of our field theory certain higher-derivative corrections will be removable via field redefinitions. We now want to show that when this happens, the corresponding corrections to hydrodynamical quantities like the current and the stress tensor are also trivial. That is, if one is not alert enough to detect the possible field redefinitions directly at the level of the Lagrangian and goes through the somewhat laborious IS procedure, at the end one is left with vanishing corrections to the hydrodynamics of the system \footnote{In the interest of full disclosure, we should mention that this happened to us for the examples discussed in sects.~\ref{firstordersuper}, \ref{firstorderfluid}.}.

To see this, suppose that the IS matching has been carried out up to $n$-th order in the derivative expansion. One then adds to the action an $n+1$-st order term:
\be
S_{n+1} [\Phi] = S_n [\Phi] + \Delta S_{n+1}[\Phi] \; .
\ee
By $\Phi$ we are collectively denoting all our fields $\phi^I$, $\psi$, and in fact the argument we are going to give applies to more general situations, e.g.~for hydrodynamical systems involving more fields. If $\Delta S_{n+1}$ is redundant, there is a field redefinition
\be \label{Phi'}
\Phi = \Phi' + G[\Phi']
\ee
such that
\be
S_n [\Phi] + \Delta S_{n+1}[\Phi] = S_n [\Phi'] + {\cal O}(\di^{n+2}) \; . 
\ee
At $n+1$-st order, we can drop the ${\cal O}(\di^{n+2})$ higher-order correction. Clearly, if one applies the IS procedure directly to the r.h.s.~of this equation, one recovers the $n$-th order results if expressed in terms of physical quantities like $\rho$, $n$, etc.---calling the fields $\Phi$ or $\Phi'$ makes no difference from this viewpoint. How does this relate to applying the IS procedure to the l.h.s.?
Recall that the starting point for the IS prescription is the stress-energy tensor and the current, and everything else follows from there. In our field theoretical framework, these are given by functional derivatives of the action with respect to the metric and to $\di_\mu \psi$, respectively:
\be \label{Tandj}
T_{\mu\nu} = -2 \frac{\delta S}{\delta g^{\mu\nu}} \; , \qquad j^\mu = \frac{\delta S}{\delta (\di_\mu \psi)} \; .
\ee
The functional derivative w.r.t.~$\di_\mu \psi$ should be evaluated by treating $\di_\mu \psi$ as a generic vector field, with arbitrary variations that vanish at the boundary. This is totally unambiguous because our action does not contain undifferentiated fields.

Now, the crucial point is that the field redefinition \eqref{Phi'} will necessarily involve derivatives. The reason is that, once plugged into $S_n$,  it is supposed to get rid of a term involving more derivatives  than those contained in $S_n$. That $G$ contains derivatives has two effects:
\begin{enumerate}
\item
It mixes the fields with their derivatives;
\item
It mixes the metric with the fields, since the fields' derivatives are implicitly contracted via the metric.
\end{enumerate}
By `mixing' here we mean a reshuffling of how the action depends on the variables involved.

As a direct consequence of item 2., the stress-energy tensor gets `contaminated' with the equations of motion---the functional derivative w.r.t.~to the metric acquires a piece proportional to the functional derivatives w.r.t~to the fields:
\begin{align}
T_{\mu\nu} & \equiv  -2 \frac{\delta S}{\delta g^{\mu\nu}}\Big |_{\Phi} \\
T_{\mu\nu} ' & \equiv  -2 \frac{\delta S}{\delta g^{\mu\nu}}\Big |_{\Phi'} = T_{\mu\nu} -2 \frac{\delta S}{\delta \Phi} \Big|_g * \frac{\delta G}{\delta g^{\mu\nu}} \Big|_{\Phi'}  \; ,
\end{align}
where the star denotes the standard integral convolution.
Therefore, the two stress-energy tensors agree {\em on-shell}.

That the same happens for the current is less trivial to see, but  equally true. Roughly speaking, it follows from item 1.~above---the functional derivative w.r.t.~$\di_\mu \psi$ acquires a piece proportional to the functional derivatives w.r.t~to the fields---but of course we cannot treat $\psi$ and $\di_\mu \psi$ as independent functions in performing functional variations, so we have to be more precise.
A crucial fact that helps us is the following. Not only does the field redefinition  \eqref{Phi'}  involve derivatives---it {\em only} involves derivatives. Meaning: the $G$ functional does not contain undifferentiated fields. If it did, it would spoil the structure of the $n$-th order Lagrangian---instead of, or on top of affecting terms of order $n+1$ and above. Recall that in our field theory all fields enter the action with at least one derivative acting on them. So, for instance, if
we redefined $\psi$ as $\psi'+ \epsilon \psi' b$ and we plugged it into $S_0 = \int F(b,y)$, we would get corrections to the Lagrangian of the form $\epsilon F' y b$ and $\epsilon F' \, \psi \, u^\mu \di_\mu b$. The first term redefines $F$. The second does not belong in our power counting scheme, because of the undifferentiated $\psi$. Perhaps more to the point: in order for the field redefinition \eqref{Phi'} to get rid of $\Delta S_{n+1}$ starting from $S_n$, it has to obey the same symmetries as $S_n$ and $\Delta S_{n+1}$, in the sense that $\Phi'$ and $\Phi$ have to transform in the same way under these symmetries. Among these symmetries, there is shift invariance for all the fields. Therefore, $G[\Phi']$ must be shift invariant.

We can now compare the currents we would get in the $\Phi$ and $\Phi'$ representations:
\begin{align}
j^\mu & \equiv   \frac{\delta S}{\delta ( \di_\mu \psi)} \\
j ' {}^\mu & \equiv  \frac{\delta S}{\delta ( \di_\mu \psi')} = j^\mu + \frac{\delta S}{\delta (\di_\alpha \Phi)} * \frac{\delta (\di_\alpha G)}{\delta ( \di_\mu \psi')}  \; .
\end{align}
The fact that $G$ only involves derivatives of the fields allows us to pull the $\di_\alpha$ out of the last functional derivative. We can then integrate it by parts (recall that the $*$ denotes a convolution), and finally use the fact that for a shift-invariant theory, the equations of motion are just (minus) the divergence of the corresponding Noether currents. We thus get
\be
j ' {}^\mu  = j^\mu + \frac{\delta S}{\delta \Phi} * \frac{\delta G}{\delta ( \di_\mu \psi')}  \; .
\ee
Like for the stress tensors, the two currents coincide on-shell.
In the appendix we will carry out the IS matching for the redundant coupling \eqref{onederivativesuper}, and confirm these general results for that case. 

A careful examination of this argument shows that the result which we proved has actually nothing to do with the IS procedure. Namely, this proof 
demonstrates simply  that the on-shell energy-momentum and particle current do not change under field redefinitions. This resonates well with the well-known result that the $S$-matrix is invariant under field redefinitions \cite{Weinberg:1995mt} \footnote{The two statements are of course related. For instance, if one couples the fluid to  dynamical gravity, the rate of  graviton emission at leading order in $G_N$ is determined by the 
on-shell fluid energy-momentum.}.

A final comment is in order. Strictly speaking, for given $T_{\mu\nu}$ and $j_\mu$ one can get different results via the IS prescription for different choices of $u_\mu$. So, the findings of this section should be interpreted as saying ``there is a choice of $u^\mu$ for which the IS procedure applied to redundant couplings gives vanishing corrections''. For different choices of $u^\mu$, one gets nontrivial-looking corrections, which however just amount to the ``corrections'' one would get by boosting the $n$-th order expressions for $T_{\mu\nu}$ and $j_\mu$.

\section{Concluding remarks}
\label{conclusions}
Let us conclude by mentioning a number of possible future directions and open questions.

Clearly, the major deficiency of this formalism is that in its present form it does not allow to discuss dissipative phenomena. It appears to be possible to introduce these by allowing couplings between the fluid Goldstones and an additional soft sector, akin to what happens in holographic
fluids, where the near-horizon bulk modes are playing the role of such a sector \cite{Nickel:2010pr}. We leave this important challenge for  future work.

Apart from providing a straightforward and clean recipe for organizing the derivative expansion, the effective field theory description brings in other benefits. Its self-consistency implies constraints that are hard to impose in the conventional  hydrodynamical formalism, such as unitarity---in the form of absence of ghosts, for example. This may lead to universal inequalities restricting the fluid properties, such as the null energy condition  \cite{DGNR}, or an upper bound on  anomaly coefficients \cite{paper2}.

It is straightforward to extend this formalism to incorporate a larger number of conserved currents. A more interesting question could be to explore alternative choices of symmetries acting on the fluid Goldstones and to see which ones may lead to interesting systems that can be realized in nature (some symmetries leading to interesting systems that are very unlikely to be realized in nature have been already explored in  studies of massive gravity \cite{Dubovsky:2004sg}).

Finally, given the recent interest in hydrodynamics with anomalous charges, an obvious application of our methods would be to reproduce the associated effects via  effective field theory. Like for the chiral Lagrangian, anomalies at low energies should  be encoded in our Goldstone Lagrangian by a Wess-Zumino term. 
We initiate exploring this in a companion paper \cite{paper2}.

\noindent
{\em Acknowledgements.}
We would like to thank David Langlois, Rob Myers and especially Sergey Sibiryakov for useful discussions.
We are supported in part by the DOE under contracts DE-FG02-92-ER40699 (LH, AN)
and DE-FG02-11ER1141743 (AN), and by NASA under contract NNX10AH14G (LH, AN).
LH thanks HKU and the IAS at HKUST for hospitality. AN thanks the Laboratoire de Physique Th\'eorique at ENS
for hospitality.

\appendix
\section{Israel-Stewart matching for first-order superfluid dynamics}

Consider the first-order correction \eqref{onederivativesuper}.
Its contributions to the stress-energy tensor and to the current are
\begin{align}
\Delta T_{\mu\nu} & = - 2 G \di_{(\mu} \psi \, \di_{\nu)} X + 2 G \Box \psi \, \di_\mu \psi \di_\nu \psi  \nonumber \\
& + \eta_{\mu\nu} \Delta {\cal L} \\
\Delta j_\mu & =  G \di_\mu X-2 G \Box \psi \, \di_\mu \psi \; . 
\end{align}
Notice that the tensor structure simplifies considerably if one defines a new velocity field
\be
u^\mu  = N \Big[u_0^\mu + \frac{1}{\sqrt{-X}} \frac{G}{2 P'} \di^\mu X \Big]\; ,
\ee
where $N$ is a normalization factor:
\be
N^2 = 1 + \Delta {\cal L}/(P' X) \; .
\ee
At first order in derivatives---or in $G$---this has the effect of aligning the full $j_\mu$ with the velocity field and, simultaneously,  of diagonalizing the full $T_{\mu\nu}$:
\begin{align}
j_\mu& \equiv j^0_\mu + \Delta j_\mu  = \nonumber \\
& =  \Big( 2P' -2 G \Box \psi + \Delta{\cal L}/X\big ) \sqrt{-X} \, u_\mu + {\cal O}(\di^2 ) \\
T_{\mu\nu} & \equiv T^0_{\mu\nu} + \Delta T_{\mu\nu} \nonumber \\ 
& = \big(2 P' X -2 G X \Box\psi + 2 \Delta {\cal L} \big) u_\mu u_\nu \nonumber \\ 
& + \big( P + \Delta {\cal L} \big)\eta_{\mu\nu} + {\cal O}(\di^2 ) \; .
\end{align}
Notice also that this redefinition of the velocity field is equivalent to the field redefinition  \eqref{psi'}.

The corrections to energy- and charge-density associated with our one-derivative term are
\begin{align}
\Delta \rho & \equiv  u^\mu u^\nu \, \Delta T_{\mu\nu} = G \, \big[  \di \psi \cdot \di X - 2 \Box \psi \, X \big]  \label{Deltarho}\\
\Delta n & \equiv - u^\mu \, \Delta j_\mu = G/ \sqrt{-X} \, \big[  \di \psi \cdot \di X - 2 \Box \psi \, X \big]  \label{Deltan}
\end{align}
We should now define  the other thermodynamic variables via the zeroth order relations between $\rho$, $n$, $\mu$, and $p$, which define the equilibrium equation of state. We find it convenient to express everything in terms of the chemical potential. From eqs.~\eqref{p(mu)}, \eqref{mu}, and \eqref{rhoandn} we get
\begin{align}
p & = p_0 - 2 P' \mu_0 \Delta \mu \\
\rho & = \rho_0 - 2 (P' - 2 P'' \mu_0^2) \mu_0 \Delta \mu  \\
n & = n_0 - 2(P'   - 2  P'' \mu_0^2   )\Delta \mu 
\end{align}
A comparison with \eqref{Deltarho}, \eqref{Deltan} gives immediately
\be
\Delta \mu = -\frac {G \, \big[  \di \psi \cdot \di X - 2 \Box \psi \, X \big] }{2(P'+2P''X)\sqrt{-X}}
\ee
and therefore
\be
\Delta p = \frac {P' G \, \big[  \di \psi \cdot \di X - 2 \Box \psi \, X \big] }{P'+2P''X} \; .
\ee

Finally, we should express the full current and stress-energy tensor in terms of the corrected physical quantities defined as above. For the current we have simply
\be
j^\mu = n u^\mu
\ee
---unmodified w.r.t.~the zeroth-order one. For the stress-energy tensor:
\begin{align}
T_{\mu\nu} & = (\rho+p) u_\mu u_\nu + p \, \eta_{\nu\nu} \nonumber \\
& + P_{\mu\nu} \, 2 G\Big[\di X \cdot \di \psi \sfrac{P'' X}{2 P'' X + P'} +  X \Box \psi \sfrac{P'}{2 P'' X + P'} \Big]  \; ,
\end{align}
where $P_{\mu\nu}$ stands for the orthogonal projector $\eta_{\mu\nu} + u_\mu u_\nu$. The second line is proportional to the zeroth-order equations of motion
\be
\di^\mu \big( P' \di_\mu \psi \big) = P' \Box \psi + P'' \di^\mu X \di_\mu \psi \; ,
\ee
and thus vanishes on-shell, as predicted.


\end{document}